\documentclass[fleqn,10pt]{wlscirep}
\usepackage[utf8]{inputenc}
\usepackage[T1]{fontenc}
\title{A back-to-back diode model applied to MoS$_2$ van der Waals Schottky diodes}

\author[1]{Jeffrey A. Cloninger}
\author[1]{Raine Harris}
\author[1]{Kristine L. Haley}
\author[1]{Randy M. Sterbentz}
\author[2]{Takashi Taniguchi}
\author[3]{Kenji Watanabe}
\author[1,*]{Joshua O. Island}
\affil[1]{Department of Physics and Astronomy, University of Nevada Las Vegas, Las Vegas, NV 89154, USA}
\affil[2]{International Center for Materials Nanoarchitectonics, National Institute for Materials Science, 1-1 Namiki, Tsukuba 305-0044, Japan}
\affil[3]{Research Center for Functional Materials, National Institute for Materials Science, 1-1 Namiki, Tsukuba 305-0044, Japan}

\affil[*]{joshua.island@unlv.edu}

\begin{abstract}
The use of metal van der Waals contacts and the implicit reduction in Fermi-level pinning in contacted semiconductors has led to remarkable device optimizations. For example, using graphene as an electrical contact allows for tunable Schottky barriers in transistors and barristors. In this study, we present a double Schottky barrier model and apply it to barrier tunable all van der Waals transistors. In a molybdenum disulfide (MoS$_2$) transistor with graphene and few-layer graphene contacts, we find that the model can be applied to extract Schottky barrier heights that agree with the Schottky-Mott rule from simple two-terminal current-voltage measurements at room temperature. Furthermore, we show tunability of the Schottky barrier \textit{in-situ} using a regional contact gate. Our results show that a basic back-to-back diode model, applied to two terminal measurements, can capture the diode properties of all-van-der-Waals transistors relatively well. 
\end{abstract}
\begin{document}

\flushbottom
\maketitle
%
%
\thispagestyle{empty}

\section*{Introduction}

Introduction of the graphene barristor showed that the Schottky barrier can be directly tuned at the interface between graphene and a semiconductor\cite{yang2012graphene}. This device, which consists of graphene on the surface of silicon with a top gate, provides a stark contrast to conventional metal contacted silicon that leads to strong Fermi level pinning (FLP). FLP has significant implications for the performance of semiconductor devices. It can impact the band alignment at the semiconductor-metal or semiconductor-insulator interface, affecting device characteristics such as threshold voltage, carrier injection, and overall electronic transport properties. 

Additionally, FLP presents a significant challenge in conventional metal-contacted two-dimensional semiconductors. Metal-induced gap states (MIGS) are responsible for strong FLP in 2D materials as in conventional semiconductors\cite{sotthewes2019universal}.
Using van der Waals metals as a contact is expected to reduce MIGS and thereby lower FLP\cite{liu2016van}. For example, accurate Schottky barriers can be extracted for van der Waals materials contacted to WS2\cite{murali2021accurate} and InSe\cite{zhao2020inse}. When gold is transferred on top of various 2D materials, as opposed to evaporated on top, reasonable Schottky barriers are also extracted\cite{liu2018approaching}. 

Diode models have been applied to numerous studies to extract device characteristics. Due to strong FLP though, these characteristics typically do not agree with theory. In the past, the back-to-back diode model has been applied to nanodevices with strong FLP, resulting in extracted parameters that often do not agree with the Schottky-Mott rule. For instance, ref. \cite{chiquito2012back} reports no change in the extracted schottky barriers from a back-to-back model for metal-semiconductor-metal junctions, even for different contact metals. 

Here, we show that a straightforward back-to-back diode model can be applied to metal van der Waals contacted MoS$_2$ junctions. The model extracts barriers that are in general agreement with the Schottky-Mott rule. In few-layer graphene contacted MoS$_2$ devices, we extract barriers of 0.46 eV to 0.58 eV, varying according to the gating conditions. For a graphene contact we extract barriers of 0.37 to 0.74 eV for different contact gate voltages. Our results here show that the simple model is capable of capturing reasonable barrier values for both graphene and few-layer graphene contacts to few-layer MoS$_2$ flakes. 

\section*{Results}

\begin{figure}[ht]
\centering
\includegraphics[width=4in]{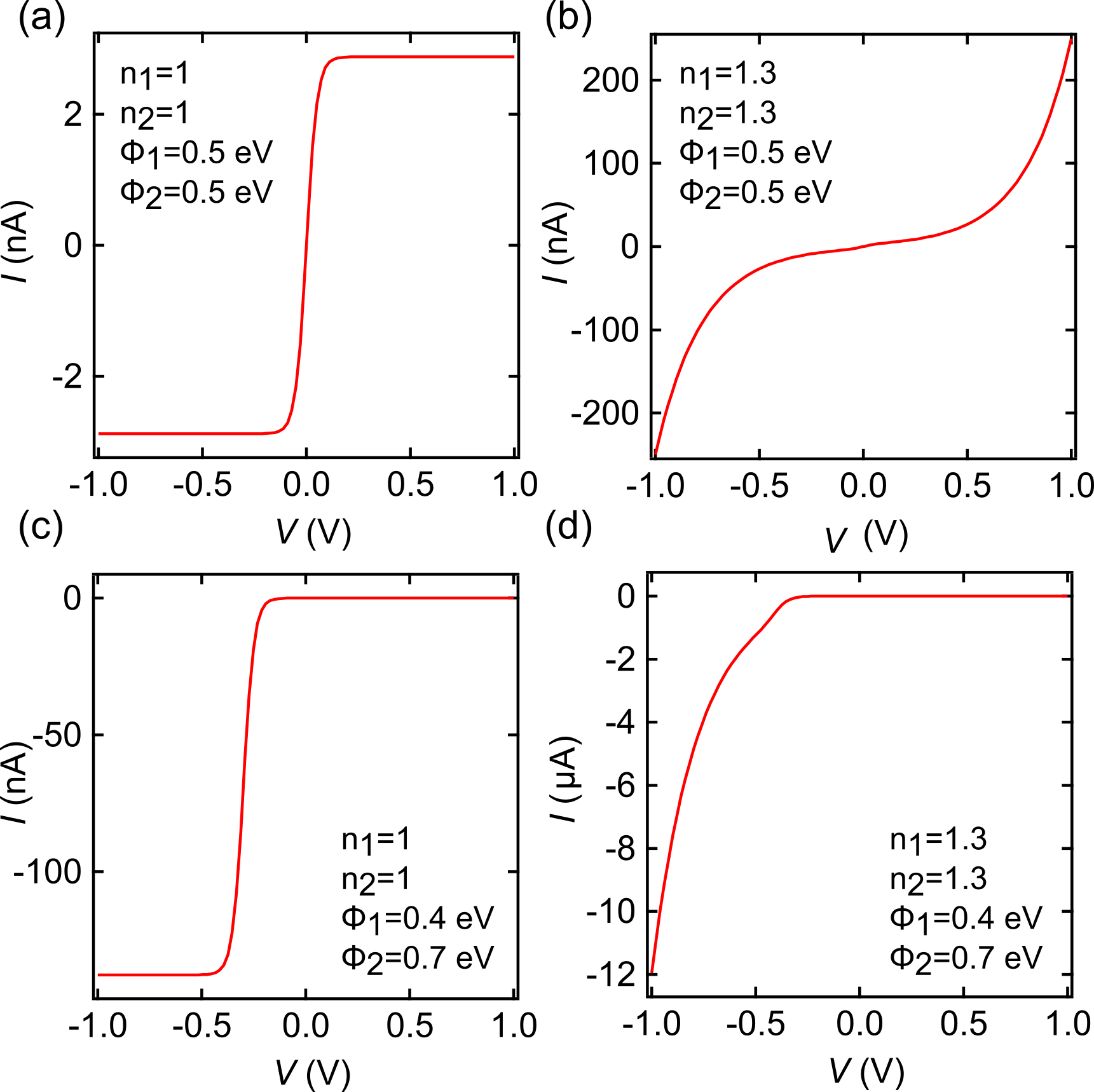}
\caption{Calculated current-voltage curves for a back-to-back diode model. (a) An ideal case in which the ideality constants are 1 and the Schottky barriers are equal. (b) A non-ideal case in which the ideality constants are 1.3 and the barriers are equal. (c) An ideal case in which the ideality constants are 1 and the barriers are not equal. (d) A non-ideal case in which the ideality constants are 1.3 and the barriers are unequal.}
\label{fig1}
\end{figure}

The back-to-back diode model has been investigated in several studies\cite{wang2020extraction, osvald2015back, grillo2021current, chiquito2012back}. Here, we primarily follow Grillo and Bartolomeo for ease of implementation\cite{grillo2021current}. Our open source python implementation of the model is available for others to use\cite{github}.
The current through each contact of the device can be written from thermionic theory as: 
\begin{equation}
    I_1=I_{s1}\left(e^{\frac{qV_1}{kT}}-1\right),I_2=-I_{s2}\left(e^{\frac{-qV_2}{kT}}-1\right),
\end{equation}
where
\begin{equation}
    I_{s1,s2}=S_{1,2}A^*T^2\left(e^{-\frac{\phi_{B01,B02}}{kT}}\right), 
\end{equation}
are the reverse saturation currents, $V_{1,2}$ are the voltage drops across each contact, $k$ is the Boltzmann constant, $T$ is the temperature, $S_{1,2}$ are the contact areas, $A^*$ is the Richardson constant, and $\phi_{B01,B02}$ are the Schottky barriers. For simplicity, we make the assumption that the voltage drops are equal across each contact, $V_1=V_2=V/2$. The total current through the device must be equal to the currents through each junction ($I_{tot}=I_1=I_2$) and the total voltage drop is equal to the sum of the voltage drops across each junction ($V=V_1+V_2$). Through a little algebra, the total current can be expressed as: 
\begin{equation}
I_{tot}=\frac{2I_{s1}I_{s2}\text{sinh}\left(\frac{qV}{2kT}\right)}{\left(I_{s1}e^{\frac{qV}{2kT}}+I_{s2}e^{-\frac{qV}{2kT}}\right)}.
\end{equation}
Figure \ref{fig1}(a,c) shows the calculated current for two ideal cases ($n_1,n_2=1$) with equal and unequal junction barriers. The exponential increase in current around zero bias is present. The current saturates at higher biases set by the reverse saturation current of the reversed-biased junction. 
For non-ideal cases in nanoscale junctions, the intrinsic Schottky barriers ($\phi_{B01,B02}$) can be replaced by effected barriers that include the voltage dependence (image charge lowering) and an ideality constant, $n$, to account for defects and interface oxides. The effective barriers can be written as:
\begin{equation}
\phi_{B1,B2}=\phi_{B01,B02}\pm eV_{1,2}\left(1-\frac{1}{n_{1,2}}\right).
\end{equation}
Two examples of non-ideal diode curves are shown in Figure \ref{fig1}(b,d) for equal and unequal barrier strengths. A more gradual nonlinearity is observed that is reminiscent of experimental current-voltage curves for nanoscale devices.  

\begin{figure}[ht]
\centering
\includegraphics[width=4in]{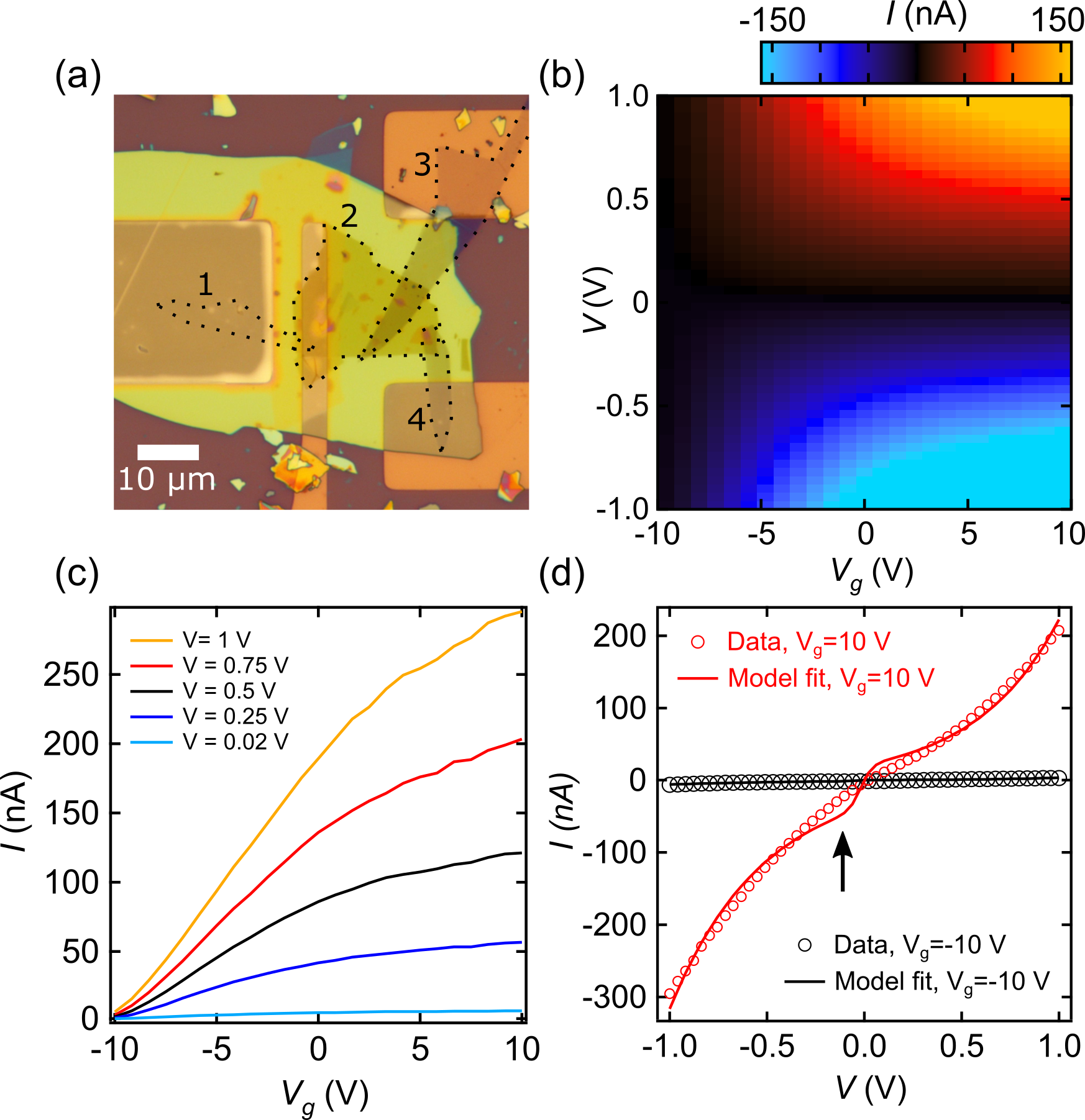}
\caption{An MoS$_2$ transistor with graphene and few-layer graphene contacts. (a) An optical image of the device. The dotted lines outline the various flakes in the van der Waals heterostructure. 1 is the graphene flake on the left side of the image, 2 is the MoS$_2$ flake, and 3 and 4 are both few-layer graphene flakes on the right side. (b) Color plot of the current through the few-layer graphene contacts (3 and 4) as a function of bias voltage ($V$) and gate voltage ($V_g$). (c) $I-V_g$ curves for increasing positive bias voltage. (d) $I-V$ curves for gate voltages of -10 V (black) and 10 V (red). The solid lines indicate fits to the $I-V$ curves using the two diode model in the main text.}
\label{fig2}
\end{figure}

This model is used to extract ideality constants ($n_1,n_2$) and Schottky barriers from simple two terminal current-voltage measurements in an all van der Waals transistor composed of MoS$_2$ with graphene and few-layer graphene contacts. Figure \ref{fig2}(a) shows an optical image of the device. To fabricate the device, prepatterned gold electrodes, accompanied by a chromium layer for adhesion, are deposited onto a heavily doped silicon wafer with 285 nm of SiO2. A van der Waals heterostructure is stacked and transferred on top of electrodes using dry stacking and heated transfer techniques\cite{wang2013one, haley2021heated}. The heterostructure is composed of MoS$_2$, few-layer graphene, graphene, and boron nitride (BN) layers used as a dielectric. On the left side, the MoS$_2$ flake (labeled 2) is connected to a gold electrode by a graphene flake (labeled 1). At the junction of the MoS$_2$ flake and the graphene, there is a gold finger gate. On the right side, the same MoS$_2$ flake is connected to two few-layer graphene flakes (labeled 3 and 4). 

We first investigate the right side of the device, using the two few-layer graphene contacts as source and drain electrodes and the highly doped silicon substrate as a back gate. Figure \ref{fig2}(b) shows a 2D color plot of the drain current as a function of bias and gate voltage. The typical transistor response is observed, with the left side of the plot indicating the OFF state and the right side of the plot the ON state. The 1D gate sweeps for increasing bias voltage are shown in Figure \ref{fig2}(c). Figure \ref{fig2}(d) shows $I-V$ curves in the OFF (black) and ON (red) states. Using the two diode model, we extract ideality constants close to 1 ($n_1=1.09$, $n_2=1.17$ in the OFF state and $n_1=1.12$, $n_2=1.14$ in the ON state) and Schottky barriers of $\phi_1=0.58$ eV, $\phi_2=0.58$ eV in the OFF state and $\phi_1=0.49$ eV, $\phi_2=0.46$ eV in the ON state. The most significant difference between the model and data is a low bias nonlinearity, highlighted by the black arrow in Figure \ref{fig2}(d). This can be attributed to the absence of the series resistance of the MoS$_2$ flake itself from the model\cite{osvald2015back}. This series resistance leads to softening of the low bias nonlinearity in devices with similar barriers for source and drain. 

The observed barrier lowering for increasingly positive gate voltage for both the source and drain is consistent with previous results showing similar behavior\cite{vaknin2020schottky}. As a result of van der Waals contact, there is minimal Fermi level pinning and the Schottky-Mott relationship, $\phi_{1,2}=\phi_m-\chi_S$ where $\phi_m$ is the work function of the metal and $\chi_S$ is the electron affinity of the semiconductor, can be directly applied. Given that the electron affinity of MoS$_2$ is 4.0 eV\cite{das2013high} and the work function of few-layer graphene (graphite) is 4.5 eV\cite{akada2019work}, we would expect a Schottky barrier of 0.5 eV. This is consistent with our results and highlights the ease of extracting barrier information from simple two-terminal measurements. 

\begin{figure}[ht]
\centering
\includegraphics[width=4in]{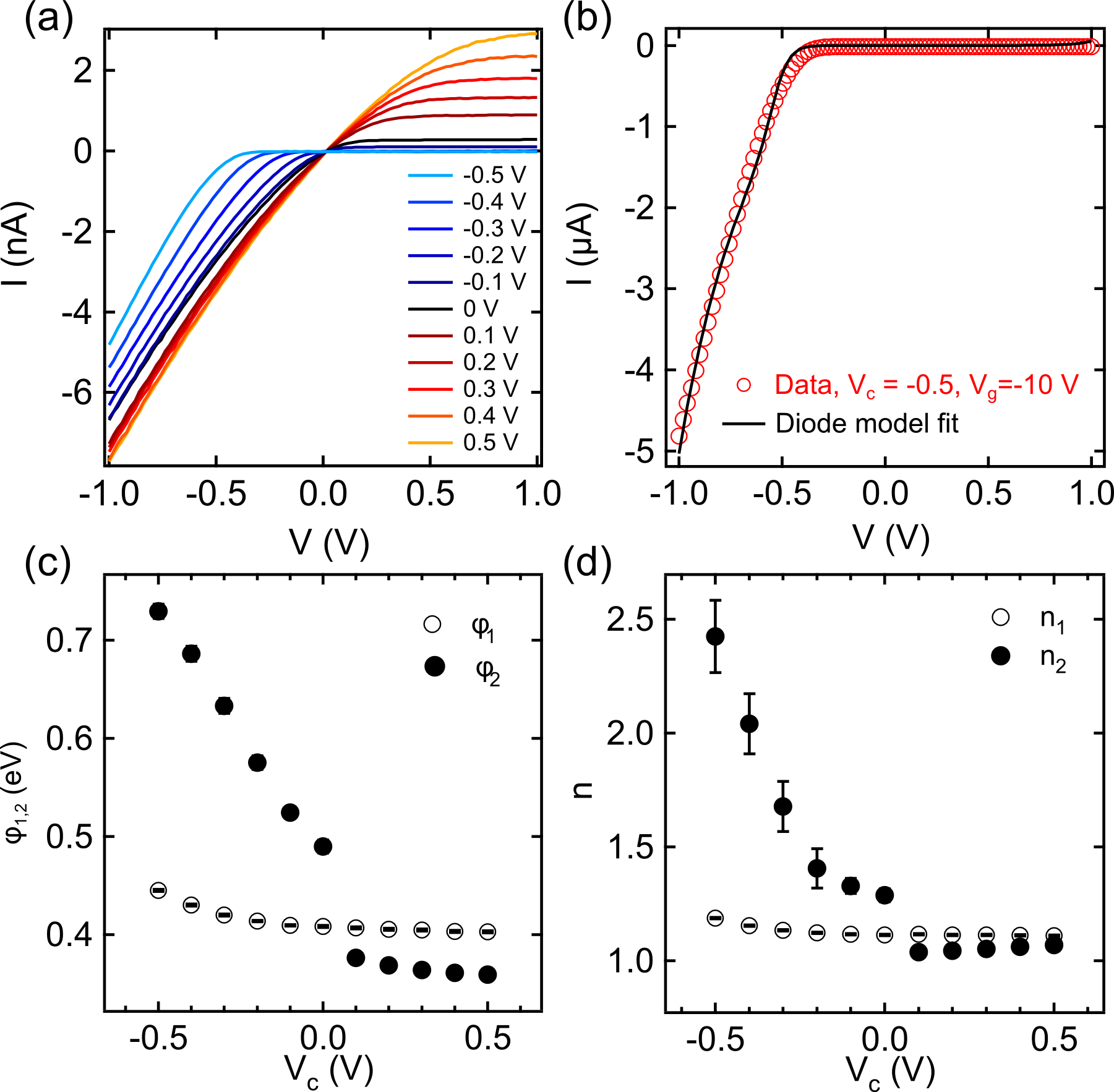}
\caption{Tunable Schottky barriers in graphene contacted MoS$_2$. (a) Current as a function of bias voltage for different contact gate voltages. (b) Current as a function of bias voltage for a contact voltage of -0.5 V (red circles). The black curve shows the fit from the diode model. (c) Extracted Schottky barriers for the graphene ($\phi_1$) and few-layer graphene contact ($\phi_2$) for the data shown in panel (a). The error bars show one-sigma uncertainty in the fitting parameters. (d) Extracted ideality factors for the graphene ($n_1$) and few-layer graphene ($n_2$) contact. The error bars show one-sigma uncertainty in the fitting parameters.}
\label{fig3}
\end{figure}

Using the contact gate, the Schottky barrier for the graphene to MoS$_2$ contact can be tuned independently. Figure \ref{fig3}(a) shows the current-voltage sweeps at $V_g=10$ V for different contact gate voltages. For negative contact gate voltages, the $I-V$ curves show a clear diode-like response, with the forward current completely suppressed. For positive contact gate voltages, this suppression is lifted and finite forward currents are observed. By employing the double diode model once again,  we can extract the Schottky barriers and ideality factors for different contact gate voltages. Figure \ref{fig3}(b) plots the curve in Figure \ref{fig3}(a) for contact gate of -0.5 V (red circles). The diode model fit (black) produces Schottky barriers of $\phi_1=0.44$ eV (few-layer graphene) and $\phi_2=0.73$ eV (graphene), and ideality factors of $n_1=1.19$ and $n_2=2.42$ (respectively). The negative contact voltage pulls the Fermi energy of graphene into the valence band and increases the Schottky barrier for the corresponding contact. The contact gate completely shields the graphene-MoS$_2$ contact from the global back-gate. 

The Schottky barriers and ideality factors from fits for all the data curves in Figure \ref{fig3} are shown in panels (c) and (d). Error bars in these figures indicate the standard deviations for each parameter from the fit. While the Schottky barrier for the few-layer graphene contact stays relatively constant over the change in contact gate voltage, the barrier at the graphene contact changes considerably, spanning more than 300 meV in energy. This behavior is also reflected in the ideality constants. 

\section*{Discussion}

Part of this modulation can be understood from the electric field-induced change to the work function of graphene. The work function of graphene is marginally lower than graphite. Both experimental and theoretical studies report the work function to be approximately 4.3 eV\cite{hibino2009dependence, leenaerts2016work, naghdi2019tuning}. This would correspond to Schottky barrier of 0.3 eV from the Schottky-Mott rule. Self-consistent density functional calculations of an explicit graphene-MoS2 heterostructure estimate the Schottky barrier to be 0.37 eV\cite{baik2017work}. This roughly aligns with our extract barrier at zero contact gate voltage of 0.48 eV. Residual doping of the graphene layer may be responsible for the small discrepancy.  
The gate-induced change in the work function of graphene has been reported using surface Kelvin probe microscopy\cite{yu2009tuning}. The Fermi energy in single-layer graphene can be calculated from\cite{yu2009tuning}:
\begin{equation}
    E_F = \text{sign}(\Delta V_g)\hbar v_F(\alpha\pi|\Delta V_g|)^{1/2}
\end{equation}
where $\alpha$ is the capacitance and $v_F$ is the Fermi velocity. A calculation of the change in the Fermi energy with gate voltage for this device is shown in the inset of Figure \ref{fig3}(d). 
For this calculation, the dielectric constant of BN is taken to be 3.76\cite{laturia2018dielectric}. The BN thickness for the graphene contact insulating layer is 6 nm, estimated by optical contrast analysis\cite{krevcmarova2019optical}. Over the 1-volt range of the contact gate, an isolated graphene layer typically exhibits a modulation of $E_F=\sim300$ meV. In the case of a graphene in contact with MoS$_2$, there is an enhanced tunability of the barrier, since the gate influences both materials. Calculations for this heterostructure show a significant change of 0.75 eV to 0.03 eV when subjected to slightly larger electric field strengths (0.01 V/Ang vs. 0.008 V/Ang) \cite{baik2017work}. This agrees with the magnitude of the Schottky barriers for negative contact gate in our measurements. 

The lack of modulation of the Schottky barrier beyond 100 mV contact gate could be due to pinning from impurity levels in the MoS$_2$ itself\cite{sachs2013doping}. Rhenium impurities are known to be present in MoS$_2$ samples and lead to impurity states near the conduction band (0.29 eV below the conduction band)\cite{sachs2013doping}. A similar trend in the change in Schottky barrier heights has been reported in graphene and few-layer graphene contacted MoS$_2$ devices\cite{qiu2015electrically,kim2020high}. The barrier height modulation flattens for increasingly positive gate voltages. 

We have shown that a straightforward back-to-back diode model can be used to extract Schottky barrier heights and ideality factors in MoS$_2$ transistors with graphene and few-layer graphene contacts. The lack of strong FLP leads to barrier heights that reasonably agree with the Schottky-Mott rule with consideration of defect doping. In an MoS$_2$ transistor with two few-layer graphene contacts, we extract barrier heights of 0.46 eV to 0.58 eV depending on gating conditions. In an MoS$_2$ transistor with a graphene contact on one side, we extract a barrier height of 0.48 eV. Using a contact gate on the graphene contacted side of the device, we showed how the barrier heights could be directly tuned. Our results support the use of a simple back-to-back diode model in extracting characteristics in van der Waals contacted 2D semiconductors. 

\section*{Methods}

Topical subheadings are allowed. Authors must ensure that their Methods section includes adequate experimental and characterization data necessary for others in the field to reproduce their work.

\section*{Acknowledgements (not compulsory)}

Acknowledgements should be brief, and should not include thanks to anonymous referees and editors, or effusive comments. Grant or contribution numbers may be acknowledged.

\section*{Author contributions statement}

Must include all authors, identified by initials, for example:
A.A. conceived the experiment(s),  A.A. and B.A. conducted the experiment(s), C.A. and D.A. analysed the results.  All authors reviewed the manuscript. 

\section*{Additional information}

To include, in this order: \textbf{Accession codes} (where applicable); \textbf{Competing interests} (mandatory statement). 

The corresponding author is responsible for submitting a \href{http://www.nature.com/srep/policies/index.html#competing}{competing interests statement} on behalf of all authors of the paper. This statement must be included in the submitted article file.

\bibliography{bib}

\end{document}